\documentclass[aoas,nameyear,seceqn,rotating,dvips]{arximspdf}
\usepackage{dcolumn}
\usepackage{graphicx}

\doi{10.1214/10-AOAS397}
\volume{5}
\issue{2A}
\pubyear{2011}
\firstpage{969}
\lastpage{993}

\makeatletter
\newcommand{\Perp}{\perp\!\!\!\!\perp}

\newcolumntype{d}[1]{D{.}{.}{#1}}
\makeatother

\begin{document}
\begin{frontmatter}

\title{Copula Gaussian graphical models and their application to
modeling functional disability data\thanksref{T1}}

\runtitle{Copula Gaussian graphical models}

\thankstext{T1}{Supported in part by NIH Grant R01 HL092071.}

\begin{aug}
\author[A]{\fnms{Adrian} \snm{Dobra}\corref{}\ead[label=e1]{adobra@uw.edu}
\ead[label=u1,url]{http://www.stat.washington.edu/adobra}}
\and
\author[B]{\fnms{Alex} \snm{Lenkoski}\ead[label=e3]{lenkoski@stat.washington.edu}}

\runauthor{A. Dobra and A. Lenkoski}
\affiliation{University of Washington and Heidelberg University}

\address[A]{Department of Statistics\\
Department of Biobehavioral Nursing\\
\quad and Health Systems\\
and\\
Center for Statistics\\
\quad and the Social Sciences\\
University of Washington\\
Box 354322\\
C-14B Padelford Hall\\
Seattle, Washington 98195-4322\\
USA\\
\printead{e1}\\
\printead{u1}
}
\address[B]{Department of Applied Mathematics\\
Heidelberg University\\
Im Neuenheimer Feld 294\\
69120 Heidelberg\\
Germany\\
\printead{e3}
}

\end{aug}

\received{\smonth{3} \syear{2009}}
\revised{\smonth{6} \syear{2010}}

%
\begin{abstract}
We propose a comprehensive Bayesian approach for graphical model
determination in observational studies that can accommodate binary,
ordinal or continuous variables simultaneously. Our new models are
called copula Gaussian graphical models (CGGMs) and embed graphical
model selection inside a semiparametric Gaussian copula. The domain of
applicability of our methods is very broad and encompasses many studies
from social science and economics. We illustrate the use of the copula
Gaussian graphical models in the analysis of a~16-dimensional
functional disability contingency table.
\end{abstract}

%
\begin{keyword}
\kwd{Bayesian inference}
\kwd{Gaussian graphical models}
\kwd{latent variable model}
\kwd{Markov chain Monte Carlo}.
\end{keyword}

\end{frontmatter}

\section{Introduction} \label{sec:intro}

The determination of conditional independence relationships through
graphical models is a key component of the statistical analysis of
observational studies. A pertinent example we will focus on in this
paper is a functional disability data set extracted from the
``analytic'' data file for the National Long Term Care Survey (NLTCS)
created by the Center of Demographic Studies at Duke University. Each
observed variable is binary and corresponds to a measure of disability
defined by an activity of daily living. This contingency table
cross-classifies information on elderly aged $65$ and above pooled
across four survey waves, 1982, 1984, 1989 and 1994---see
\citet{manton2et21993} for more details. The $16$ dimensions of this
table correspond to six activities of daily living (ADLs) and ten
instrumental activities of daily living (IADLs). Specifically, the ADLs
relate to hygiene and personal care: eating (ADL1), getting in/out of
bed (ADL2), getting around inside (ADL3), dressing (ADL4), bathing
(ADL5) and getting to the bathroom or using a toilet (ADL6). The IADLs
relate to activities needed to live without dedicated professional
care: doing heavy house work (IADL1), doing light house work (IADL2),
doing laundry (IADL3), cooking (IADL4), grocery shopping (IADL5),
getting about outside (IADL6), travelling (IADL7), managing money
(IADL8), taking medicine (IADL9) and telephoning (IADL10). For each
ADL/IADL measure, subjects were classified as being either healthy
(level 1) or disabled (level 2) on that measure. The methodology we
develop in this paper allows us to determine the complex pattern of
conditional associations that exist among the $16$ daily living
activities. This represents a critical issue that was left unexplored
in previous analyses of this data set
[\citet{erosheva2008}; \citet{fien1hers1rina1zhou12008}].

In fact, the domain of applicability of our methods is not restricted
to contingency tables. Since multivariate data sets arising from social
science or economics typically contain variables of many types, our
goal is to develop an approach to graphical model determination that is
broad enough to be applicable to any study that involves a mixture of
binary, ordinal and continuous variables.

Most of the research efforts in the graphical models literature have
been focused on multivariate normal models or on log-linear models---see, for example, the monographs of \citet{lauritzen21996} and
\citet{whittaker1990}. These models relate to data sets that contain
exclusively continuous or categorical variables. CG distributions
[\citet{lauritzen21996}] constitute the basis of a class of graphical
models for mixed variables, but they impose an overly restrictive
assumption: the conditional distribution of the continuous variables
given the discrete variables must be multivariate normal. As such, the
three main classes of graphical models are too restrictive to be widely
applicable to social science or economics studies.

Copulas [\citet{nelsen21999}] provide the theoretical framework in which
multivariate associations can be modeled separately from the univariate
distributions of the observed variables. \citet{genest2neslehova22007}
advocate the use of copulas when modeling multivariate distributions
involving discrete variables. In this paper we employ the Gaussian
copula and further require conditional independence constraints on the
inverse of its correlation matrix. The resulting models are called
copula Gaussian graphical models (CGGMs) because they only impose a
multivariate normal assumption for a~set of latent variables which are
in a one-to-one correspondence with the set of observed variables. A
related approach for inference in Gaussian copulas has been developed
by \citet{pitt2et22006}. Their framework involves parametric models for
Gaussian copulas and the univariate marginal distributions of the
observed variables.We treat these marginal distributions as nuisance
parameters and focus on the determination of graphical models.

The structure of the paper is as follows. In Section
\ref{sec:bayeslearn} we formally introduce Gaussian graphical models
(GGMs) and describe a Bayesian framework for inference in this class of
models. In Section \ref{sec:ordcar} we discuss modeling aspects related
to binary and ordinal variables. In Section \ref{sec:copulaggm} we show
how to extend GGMs to represent conditional independence associations
in a latent variables space. We also present a Bayesian model averaging
approach for graph identification and estimation in CGGMs. In Section
\ref{sec:examples} we analyze the NLTCS functional disability data
together with another six-dimensional contingency table using CGGMs. We
discuss our proposed methodology in Section \ref{sec:discussion}.

\section{Gaussian graphical models} \label{sec:bayeslearn}

We let $X=X_V$, $V=\{1,2,\ldots,p\}$, be a~random vector with a joint
distribution $p(X_V)$. The conditional independence relationships among
$\{ X_v\dvtx v\in V\}$ under $p(X_V)$ can be summarized in a graph
$G=(V,E)$, where each vertex $v\in V$ corresponds with a random
variable $X_v$ and $E\subset V\times V$ are undirected edges
[\citet{whittaker1990}]. Here ``undirected'' means that $(v_1,v_2)\in E$
is equivalent with $(v_2,v_1)\in E$.

The absence of an edge between $X_{v_1}$ and $X_{v_2}$ corresponds with
the conditional independence of these two random variables given the
remaining variables under $p(X_V)$ and is denoted by
%
\begin{eqnarray} \label{eq:pairwisemarkov}
X_{v_1}\Perp X_{v_2}\mid X_{V\setminus\{ v_1,v_2\}}.
\end{eqnarray}
This is called the pairwise Markov property relative to $G$, which in
turn implies the local as well as the global Markov properties relative
to $G$ [\citet{lauritzen21996}].

We denote by $\mathcal{G}_V$ the set of all $2^{p(p-1)/2}$ undirected
graphs with vertices~$V$. Since $\mathcal{G}_V$ contains many graphs
even for relatively small values of $p$, it cannot be enumerated and
has to be visited using stochastic search methods
[\citet{madigan2york21995}; \citet{jones2et22005}; \citet{lenkoski2dobra22008}]. Such
algorithms move through $\mathcal{G}_V$ using neighborhood sets
$\operatorname{nbd}(G)\subset\mathcal{G}_V$ for $G\in\mathcal{G}_V$. The
neighborhood of a graph $G\in\mathcal{G}_V$ is comprised of all the
graphs obtained from $G$ by adding or deleting one edge. These
neighborhood sets are symmetric and link any two graphs through a path
of graphs such that two consecutive graphs on this path are neighbors
of each other. We remark that the neighborhood sets associated with~$\mathcal{G}_V$ contain the same number of graphs $p(p-1)/2$.

Furthermore, we assume that $X=X_V$ follows a $p$-dimensional
\mbox{multivariate} normal distribution $\mbox{N}_p(0,K^{-1})$ with precision
matrix $K = (K_{v_1,v_2})_{1\le v_1,v_2 \le p}$. We let $x^{(1:n)}=(
x^{(1)},\ldots,x^{(n)})^{T}$ be the observed data of $n$ independent
samples of $X$. The likelihood function is proportional to
%
\begin{eqnarray} \label{eq:likelihood}
p\bigl(x^{(1:n)}|K\bigr) \propto(\operatorname{det} K)^{n/2}\exp\bigl\{-\tfrac{1}{2}\langle
K,U \rangle\bigr\},
\end{eqnarray}
where $U = \sum_{j=1}^nx^{(j)}x^{(j)T}$, and $\langle A,B\rangle=
\operatorname{tr}(A^{T}B)$ denotes the trace inner product. We assume that the
data have been centered and scaled, so that the sample mean of each
$X_v$ is zero and its sample variance is one.

A graphical model $G=(V,E)$ for $N_p(0,K^{-1})$ is called a Gaussian
graphical model (GGM) and is constructed by constraining some of the
off-diagonal elements of $K$ to zero. For example, the pairwise Markov
property (\ref{eq:pairwisemarkov}) holds if and only if
$K_{v_1,v_2}=0$. This implies that the edges of $G$ correspond with
the off-diagonal nonzero elements of $K$, that is,
\mbox{$E=\{(v_1,v_2)\mid K_{v_1,v_2}\ne0, v_{1}\ne v_{2}\}$}. Given $G$, the
precision matrix $K$ is constrained to the cone $P_G$ of symmetric
positive definite matrices with entries $K_{v_1,v_2}$ equal to zero for
all $(v_1,v_2)\notin E$, $v_{1}\ne v_{2}$.

We consider a $G$-Wishart prior $\mbox{W}_G(\delta,D)$ for $K$ with
density
%
\begin{equation} \label{eq:wishart}
p(K|G) =\frac{1}{I_G(\delta,D)}(\operatorname{det}
K)^{(\delta-2)/2}\exp\biggl\{-\frac{1}{2}\langle K,D\rangle\biggr\},
\end{equation}
with respect to the Lebesgue measure on $P_G$
[\citet{roverato22002}; \citet{atay4kayis2massam22005}; \citet{letac2massam22007}]. The
normalizing cons\-tant~$I_G(\delta,D)$ is finite provided $\delta>2$ and
$D$ is positive definite [\citet{diaconis2ylvisaker21979}]. If $G$ is
the complete graph with $p$ vertices (i.e., there are no missing
edges), $W_G(\delta,D)$ reduces to the Wishart distribution
$W_p(\delta,D)$, hence, its normalizing constant is
%
\begin{equation} \label{eq:normconstcomplete}
I_G(\delta,D)  =  2^{(\delta+p-1)p/2}
\Gamma_{p}\{(\delta+p-1)/2\}(\operatorname{det} D)^{-(\delta+p-1)/2},
\end{equation}
where
$\Gamma_{p}(a)=\pi^{p(p-1)/4}\prod_{i=0}^{p-1}\Gamma(a-\frac{i}{2})$
for $a>(p-1)/2$ [\citet{muirhead22005}]. If $G$ is decomposable,
$I_G(\delta,D)$ is explicitly calculated [\citet{roverato22002}]. For
nondecomposable graphs, the Monte Carlo method of
\citet{atay4kayis2massam22005} can be used to numerically approximate
$I_G(\delta,D)$ in a fast and accurate manner.

Throughout this paper we set the prior parameters for $K$ to $\delta=3$
and $D=I_p$, the $p$-dimensional identity matrix. From equations
(\ref{eq:likelihood}) and (\ref{eq:wishart}) we see that the
interpretation of this prior is that the components of $X$ are
independent apriori and that the ``weight'' of the prior is equivalent
to one observed sample.

The $G$-Wishart prior is conjugate to the likelihood
(\ref{eq:likelihood}), thus, the posterior distribution of $K$ given
$G$ is $\mbox{W}_G(\delta+n,D+U)$, that is,
\[
p\bigl( K|x^{(1:n)},G\bigr) = \frac{1}{I_G(\delta+n,D+U)}(\operatorname{det}
K)^{(\delta+n-2)/2}\exp\biggl\{-\frac{1}{2}\langle K,D+U\rangle\biggr\}.
\]
Given $K\in P_G$, the regression of $X_v$ on the remaining elements of
$X$ depends only on the neighbors of $v$ in $G$:
%
\begin{equation} \label{eq:graphreg}
p\bigl(X_v|X_{V\setminus\{v\}}=x_{V\setminus\{v\}},K\bigr) = \mathrm{N}\biggl(-\sum
_{v'\in bd_G(v)}\frac{K_{v,v'}}{K_{v,v}}x_{v'}, \frac{1}{K_{v,v}}\biggr),
\end{equation}
where $bd_G(v) = \{v'\in V\dvtx (v,v')\in E\}$.

The Cholesky decomposition of a matrix $K\in P_{G}$ is
$K=\phi^{T}\phi$, where $\phi$ is an upper triangular matrix with
$\phi_{v,v}>0$, $v\in V$. \citet{roverato22002} proved that the set
$\nu(G)$ of the free elements of $\phi$ consists of the diagonal
elements together with the elements that correspond with the edges of
$G$, that is,
\[
\nu(G) = \{ (v_{1},v_{1})\dvtx v_{1}\in V\} \cup\{
(v_{1},v_{2})\dvtx v_{1}<v_{2}\mbox{ and } (v_{1},v_{2})\in E\}.
\]
Once the free elements of $\phi$ are known, the remaining elements are
also known. More specifically, we have $\phi_{1,v_{2}}=0$ if $v_{2}\ge
2$ and $(1,v_{2})\notin E$. We also have
\[
\phi_{v_{1},v_{2}} = -\frac{1}{\phi_{v_{1},v_{1}}}\sum
_{v=1}^{v_{1}-1}\phi_{v,v_{1}}\phi_{v,v_{2}}
\]
for $2\le v_{1}<v_{2}$ and $(v_{1},v_{2})\notin E$. The determination
of the elements of $\phi$ that are not free based on the elements of
$\phi$ that are free is called the completion of $\phi$ with respect to
$G$ [\citet{roverato22002}; \citet{atay4kayis2massam22005}]. It is useful to
remark that the free elements of $\phi$ fully determine the matrix $K$.
The development of our framework involves the Jacobian of the
transformation that maps $K\in P_{G}$ to the free elements of $\phi$
[\citet{roverato22002}]:
\[
J(K \rightarrow\phi) = 2^{p}\prod_{v=1}^{p}\phi_{v,v}^{d^{G}_{v}+1},
\]
where $d^{G}_{v}$ is the number of elements in $bd_G(v)\cap
\{v+1,\ldots,p\}$.

\section{Incorporating binary and ordinal categorical variables}
\label{sec:ordcar}

A varia\-ble~$X_{v}$ that takes a finite number of ordinal values
$\{1,2,\ldots,d_v\}$, with $d_v\ge2$, is incorporated in our modeling
framework by introducing a continuous latent variable $Z_{v}$
underlying $X_{v}$---see, for example, \citet{muthen1984}. We denote
by $\{x^{(1)}_{v},\ldots,x^{(n)}_{v}\}$ the observed samples associated
with $X_{v}$. The samples from $Z_{v}$ are denoted by
$\{z^{(1)}_{v},\ldots,z^{(n)}_{v}\}$. Typically the relationship
between $X_{v}$ and its surrogate $Z_{v}$ is expressed through some
thresholds $\tau_v=(\tau_{v,0},\tau_{v,1},\ldots,\tau_{v,w_v})$ with
$-\infty=\tau_{v,0}<\tau_{v,1}<\cdots<\tau_{v,w_v}=\infty$. Formally,
we set [\citet{dunson2006}]
%
\begin{equation} \label{eq:linkfct}
x^{(j)}_{v}=\sum_{l=1}^{w_v}l\times\mathbf{1}_{\{ \tau
_{v,l-1}<z^{(j)}_{v}\le\tau_{v,l}\}},\qquad j=1,2,\ldots,n.
\end{equation}
This model is identifiable if the value of $\tau_{v,1}$ is fixed at a
certain value. We follow an idea originally suggested by
\citet{hoff22007} that does not explicitly involve the thresholds
$\tau_v$. This approach is based on the remark that the relationship
between the observed and latent samples satisfies the constraints
%
\begin{equation} \label{eq:latentconstraints}
\quad\quad x^{(j_{1})}_{v}<x^{(j_{2})}_{v}\quad\Rightarrow\quad
z^{(j_{1})}_{v}<z^{(j_{2})}_{v},\qquad
z^{(j_{1})}_{v}<z^{(j_{2})}_{v}\quad\Rightarrow\quad x^{(j_{1})}_{v}\le x^{(j_{2})}_{v}
\end{equation}
for $1\le j_1\ne j_2\le n$. We see that if $X_{v}$ and $Z_{v}$ are
related as in (\ref{eq:linkfct}), then (\ref{eq:latentconstraints})
holds. If (\ref{eq:latentconstraints}) holds, then (\ref{eq:linkfct})
also holds by choosing $\tau_{v,l}=\max\{z^{(j)}_{v}\dvtx x^{(j)}_{v}=l\}$
for $l=1,\ldots,w_v-1$. It follows that, given the observed data
$x^{(1:n)}$, the latent samples
$z^{(1:n)}=(z^{(1)},z^{(2)},\ldots,z^{(n)})$ are constrained to belong
to the set\looseness=-1
\[
A\bigl(x^{(1:n)}\bigr) = \bigl\{ z^{(1:n)}\in
{\mathbf{R}}^{n\times p}\dvtx
L^{j}_{v}\bigl(z^{(1:n)}\bigr)<z^{(j)}_{v}<U^{j}_{v}\bigl(z^{(1:n)}\bigr)\bigr\},
\]
where
\begin{eqnarray} \label{eq:bounds}
L^{j}_{v}\bigl(z^{(1:n)}\bigr)
&=&
\max\bigl\{ z^{(k)}_{v}\dvtx x^{(k)}_{v}<x^{(j)}_v\bigr\},\nonumber
\\[-8pt]\\[-8pt]
U^{j}_{v}\bigl(z^{(1:n)}\bigr)
&=&
\min\bigl\{ z^{(k)}_{v}\dvtx x^{(j)}_{v}<x^{(k)}_v\bigr\}.\nonumber
\end{eqnarray}
If the value $x^{(j)}_v$ is missing from the observed data, we define
$L^{j}_{v}(z^{(1:n)})=-\infty$ and $U^{j}_{v}(z^{(1:n)})=\infty$.

\section{Copula Gaussian graphical models} \label{sec:copulaggm}

We assume that an observed variable~$X_{v}$ can be binary, categorical
with ordered categories, count or continuous. We denote by $F_{v}$ the
univariate distribution of $X_{v}$ and by $F_v^{-1}$ the pseudo-inverse
of~$F_v$. Given a precision matrix $K$, we model the joint distribution
of $X=X_{V}$ as follows [see also \citet{hoff22007}]:
%
\begin{eqnarray}\label{eq:relationship}
Z_V
& \sim&
N_p(0,K^{-1}),\nonumber
\\
\widetilde{Z}_v
& = &
Z_v/(K^{-1})_{v,v}^{1/2},\qquad v\in V,
\\
X_v
& = &
F_v^{-1}(\Phi(\widetilde{Z}_v)),\qquad v\in V.\nonumber
\end{eqnarray}
In (\ref{eq:relationship}) the joint distribution of the latent
variables is multivariate normal $\widetilde{Z}=\widetilde{Z}_V\sim
N_p(0,\Upsilon(K))$, where $\Upsilon(K)$ is a correlation matrix with
entries
%
\begin{equation} \label{eq:cork}
\Upsilon_{v_1,v_2}(K)=\frac{(K^{-1})_{v_1,v_2}}{\sqrt
{(K^{-1})_{v_1,v_1}(K^{-1})_{v_2,v_2}}}.
\end{equation}
The joint distribution $F$ of $X=X_V$ is subsequently a function of the
correlation matrix $\Upsilon(K)$ and the univariate distributions $F_v$
of $X_v$:
\begin{eqnarray*}
p(X_1\le x_1,\ldots,X_p\le x_p)
& = &
F(x_1,\ldots,x_p|\Upsilon(K),F_1,\ldots,F_p),
\\
& = &
C(F_1(x_1),\ldots,F_p(x_p)|\Upsilon(K)),
\end{eqnarray*}
where
%
\begin{eqnarray} \label{eq:gausscopula}
\quad\quad C(u_1,\ldots,u_p|\Upsilon^{\prime}) = \Phi_p(\Phi^{-1}(u_1),\ldots
,\Phi^{-1}(u_p)|\Upsilon^{\prime})\dvtx [0,1]^p\rightarrow[0,1]
\end{eqnarray}
is the Gaussian copula with $p\times p$ correlation matrix
$\Upsilon^{\prime}$ [\citet{nelsen21999}]. Here~$\Phi(\cdot)$ represents
the CDF of the standard normal distribution and
$\Phi_p(\cdot|\Upsilon)$ is the CDF of $N_p(0,\Upsilon)$.

We avoid the need to formally make assumptions regarding the parametric
representation of $\{ F_v\dvtx v\in V\}$, which could be a daunting task for
most real world data sets, by treating their marginal distributions as
nuisance parameters. Moreover, we reduce our model parameters to the
correlation matrix of the Gaussian copula (\ref{eq:gausscopula}). This
means that we focus on the joint distribution of the latent variables
$\widetilde{Z}_V$ whose relationships with the observed variables
$X_{V}$ are given by (\ref{eq:relationship}). Since $F_v^{-1}(\cdot)$
and $\Phi(\cdot)$ are nondecreasing, (\ref{eq:relationship}) implies
(\ref{eq:latentconstraints}) which does not depend on the marginal
distributions $\{ F_v\dvtx v\in V\}$. The converse is also true: if the
relationship (\ref{eq:latentconstraints}) between the observed and
latent samples holds, then (\ref{eq:relationship}) also holds by
replacing $F_{v}$ with the empirical distribution of $X_{v}$.

As suggested by \citet{hoff22007}, inference in the latent variables
space can be performed by substituting the observed data $x^{(1:n)}$
with the event $\mathcal{D} = \{ z^{(1:n)}\in A(x^{(1:n)})\}$. We write
the likelihood function as
\[
p\bigl( x^{(1:n)}|K, \{ F_{v}\dvtx v\in V\}\bigr)
=
p( \mathcal{D} | K) p\bigl(x^{(1:n)}|\mathcal{D},\Upsilon(K),\{ F_{v}\dvtx v\in V\}\bigr).
\]

In this decomposition $p( \mathcal{D} | K)$ is the only part of the
observed data likelihood that is relevant for making inference on $K$.
Furthermore, $p( \mathcal{D} | K)$ does not depend on $\{ F_v\dvtx v\in
V\}$. \citet{hoff22007} calls $p( \mathcal{D} | K)$ the extended rank
likelihood and constructs a Gibbs sampler with stationary distribution
%
\begin{equation} \label{eq:hoff}
p( K | \mathcal{D})  \propto p( \mathcal{D} | K) p( K),
\end{equation}
where $K$ follows a Wishart prior distribution $W_{p}(\delta,D)$.

We are interested in modeling the conditional independence
relationships among the latent variables $Z=Z_{V}$ using Gaussian
graphical models. We go one step further compared to \citet{hoff22007}
and impose zero constraints in the precision matrix $K$ according to a
graph $G$. We refer to the graphical models constructed in the latent
space as copula Gaussian graphical models (CGGMs). The inference
approach described in \citet{hoff22007} is equivalent to reducing the
set of candidate graphs to only one graph. This graph is the full graph
in which all the edges are present and none of the off-diagonal
elements of $K$ are constrained to zero.

The Markov properties associated with a CGGM are guaranteed to
translate into Markov properties for the observed variables if all the
marginals $\{ F_v\dvtx v\in V\}$ are continuous
[\citet{liu4lafferty4wasserman42009}]. The presence of some discrete
observed variables might induce additional dependencies among the $X$'s
that are not modeled in a CGGM, but such dependencies can be regarded
as having a secondary relevance since they emerge from the marginals
$\{ F_v\dvtx v\in V\}$. The conditional independence graphs for the latent
variables could contain edges then that do not necessarily correspond
with conditional independence relationships in the observed variables
space. Conversely, there might exist conditional independence
relationships among the observed variables that are not represented in
conditional independence graphs that involve latent variables.

\subsection{Bayesian inference in copula Gaussian graphical models}
\label{sec:mcmc}

Let \mbox{$G\in\mathcal{G}_V$} be a graph defining a CGGM. The joint
posterior distribution of $K\in P_G$ and the graph $G$ is given by
%
\begin{equation} \label{eq:joint}
p(K,G|\mathcal{D}) \propto p(\mathcal{D}| K)p(K|G)p(G).
\end{equation}
The prior distribution of $K$ conditional on $G$ is $G$-Wishart
$W_G(\delta,D)$ and the prior distribution over $\mathcal{G}_V$ is
uniform, that is, $p(G)\propto1$. Other choices of priors on the graphs
space $\mathcal{G}_V$ take into consideration the implied distribution
on the number of edges [\citet{wong2et22003}], encourage sparsity
[\citet{jones2et22005}] or have multiple testing correction properties
[\citet{scott2berger22006}].

We describe a Markov chain Monte Carlo sampler for the joint
distribution~(\ref{eq:joint}). We consider two strictly positive
precision parameters $\sigma_{p}$ and $\sigma_{g}$ that remain fixed
throughout at some small values, for example,
$\sigma_{p}=\sigma_{g}=0.1$. Given the current state of the chain
$(K^{s},G^{s})$, its next state $(K^{s+1},G^{s+1})$ is generated by
sequentially performing the following updates.

\textit{Step} 1: \textit{Resample the latent data}. For each $v\in V$
and $j\in \{1,2,\ldots,n\}$, we update the latent value $z^{(j)}_v$ by
sampling from its full conditional distribution. The distribution of
$Z_v$ conditional on $Z_{V\setminus\{v\}}=z^{(j)}_{V\setminus\{v\}}$ is
$N( \mu_{v},\sigma^{2}_{v})$ truncated to the interval $[L^j_v,U^j_v]$,
where $\mu_v = -\sum_{v'\in\mathrm{bd}_G(v)}\frac
{K^{s}_{v,v'}}{K^{s}_{v,v}}z^{(j)}_{v'}$ and $\sigma^2_v=\frac
{1}{K^{s}_{v,v}}$---see (\ref{eq:graphreg}). The bounds $L^j_v$ and
$U^j_v$ are given in (\ref{eq:bounds}). The new value of $z^{(j)}_v$ is
obtained by sampling from this truncated normal distribution.

\textit{Step} 2: \textit{Resample the precision matrix}. We
sequentially perturb the free elements $\{ \phi^{s}_{v_{1},v_{2}}\dvtx
(v_{1},v_{2})\in \nu(G^{s})\}$ in the Cholesky decomposition
$K^{s}=(\phi^{s})^{T}\phi^{s}$ around their current value. Here
$\phi^{s}$ is upper triangular. We perform a~Metro\-polis--Hastings
update of $K^{s}$ associated with a diagonal element
$\phi^{s}_{v_{1},v_{1}}>0$ by sampling a value $\gamma$ from a
$N(\phi^{s}_{v_{1},v_{1}},\sigma^{2}_{p})$ distribution truncated below
at~$0$, that is,
\[
\gamma\sim q(u|\phi^{s}_{v_{1},v_{1}}) \propto\frac{1}{\sigma
_{p}\Phi(\phi^{s}_{v_{1},v_{1}}/\sigma_{p})}\exp\biggl( -\frac{(u-\phi
^{s}_{v_{1},v_{1}})^{2}}{2\sigma_{p}^{2}}\biggr).
\]
We take $K^{\prime}=(\phi^{\prime})^{T}\phi^{\prime}$, where
$\phi^{\prime}$ is such that its free elements coincide with the free
elements of $\phi^{s}$, with the exception of the $(v_{1},v_{1})$
element which is set to $\gamma$. The elements of $\phi^{\prime}$ that
are not free are obtained by the completion operation described in
Section \ref{sec:bayeslearn}. The acceptance probability of the update
of $K^{s}$ to $K^{\prime}$ is $\min\{ R_{p},1\}$, where
\begin{eqnarray*}
R_{p}
& = &
\frac{p(K^{\prime}|z^{(1:n)},G^{s})}{p(K^{s}|z^{(1:n)},G^{s})}\frac{J(K^{\prime}\rightarrow\phi^{\prime})}{J(K^{s} \rightarrow\phi^{s})}\frac
{q(\phi^{s}_{v_{1},v_{1}}|\gamma)}{q(\gamma|\phi ^{s}_{v_{1},v_{1}})},
\\
& = &
\frac{\Phi(\phi^{s}_{v_{1},v_{1}}/\sigma_{p})}{\Phi(\gamma/\sigma_{p})}\biggl( \frac{\gamma}{\phi^{s}_{v_{1},v_{1}}}\biggr)^{\delta+n+d^{G^{s}}_{v_{1}}-1} R_{p}^{\prime}.
\end{eqnarray*}
Here we denote
\[
R_{p}^{\prime} = \exp\Biggl\{-\frac{1}{2}\Biggl\langle K^{\prime}-K^{s},D+\sum
_{j=1}^n z^{(j)}z^{(j)T}\Biggr\rangle\Biggr\}.
\]

Next we consider a free off-diagonal element $\phi^{s}_{v_{1},v_{2}}$,
where $v_{1}<v_{2}$ and $(v_{1},v_{2})\in\nu(G^{s})$. We sample a
candidate value $\gamma^{\prime}$ from a
$N(\phi^{s}_{v_{1},v_{2}},\sigma^{2}_{p})$ distribution. As before, we
take $K^{\prime}=(\phi^{\prime})^{T}\phi^{\prime}$, where
$\phi^{\prime}$ and $\phi^{s}$ have the same free elements with the
exception of the $(v_{1},v_{2})$ element that has
$\phi^{\prime}_{v_{1}v_{2}}=\gamma^{\prime}$. The remaining nonfree
elements of $\phi^{\prime}$ are obtained through completion. Due to the
symmetry of the proposal distribution and the fact that $\operatorname{det}
K^{s}=\prod_{v=1}^{p}(\phi^{s}_{v,v})^{2}=\prod_{v=1}^{p}(\phi
^{\prime}_{v,v})^{2}=\operatorname{det}
K^{\prime}$, the candidate matrix $K^{\prime}$ is accepted with
probability $\min\{R_{p}^{\prime},1\}$.

Since $K^{s}\in P_{G^{s}}$, the candidate matrix $K^{\prime}$
associated with each free element in $\nu(G^{s})$ must also belong to
$P_{G^{s}}$. The precision matrix that is obtained after performing all
the Metropolis--Hastings updates is $K^{s+1/2}\in P_{G^{s}}$.
$ $

\textit{Step} 3: \textit{Resample the graph}. We consider the Cholesky
decomposition $K^{s+1/2}=(\phi^{s+1/2})^{T}\phi^{s+1/2}$ where
$\phi^{s+1/2}$ is upper triangular. We randomly choose a pair
$(v_{1},v_{2})$, $v_{1}<v_{2}$. If there is no edge between $v_{1}$ and
$v_{2}$ in~$G^{s}$, that is, $(v_{1},v_{2})\notin\nu(G^{s})$, we add
this edge to $G^{s}$ to obtain a candidate graph~$G^{\prime}$. This
implies $bd_{G^{\prime}}(v_{1}) = bd_{G^{s}}(v_{1})\cup\{ v_{2}\}$,
hence, $d_{v_{1}}^{G^{\prime}}=d_{v_{1}}^{G^{s}}+1$. Moreover,
$\nu(G^{\prime})=\nu(G^{s})\cup\{(v_{1},v_{2})\}$. We define an upper
diagonal matrix~$\phi^{\prime}$ such that
$\phi^{\prime}_{v_{1}^{\prime},v_{2}^{\prime}}=\phi
^{s+1/2}_{v_{1}^{\prime},v_{2}^{\prime}}$
for all $(v^{\prime}_{1},v^{\prime}_{2})\in\nu(G^{s})$. The value of
$\phi^{\prime}_{v_{1},v_{2}}$ is set by sampling from a
$N(\phi^{s+1/2}_{v_{1},v_{2}},\sigma_{g}^{2})$ distribution. The
remaining elements of~$\phi^{\prime}$ are determined through completion
with respect to the graph $G^{\prime}$. We see that~$\phi^{\prime}$ has
one additional free element with respect to $\phi^{s+1/2}$ whose value
was randomly chosen by perturbing the nonfree $(v_{1},v_{2})$ element
of $\phi^{s+1/2}$.

We take the candidate precision matrix
$K^{\prime}=(\phi^{\prime})^{T}\phi^{\prime}\in P_{G^{\prime}}$. Since
the dimensionality of the parameter space increases by one, we must
make use of the reversible jump Markov chains methodology proposed by
\citet{green31995}. We accept the update of $(K^{s+1/2},G^{s})$ to
$(K^{\prime},G^{\prime})$ with probability $\min\{R_{g},1\}$, where
$R_{g}$ is given by
\begin{eqnarray*}
&&
\frac{p(z^{(1:n)}|K^{\prime})p(K^{\prime}|G^{\prime
})}{p(z^{(1:n)}|K^{s+1/2})p(K^{s+1/2}|G^{s})}\frac{|\operatorname{nbd}(G^{s})|}{|\operatorname{nbd}(G^{\prime})|}
\\
&&\qquad{}\times
\frac{J(K^{\prime} \rightarrow\phi^{\prime})}{J(K^{s+1/2} \rightarrow\phi^{s+1/2})}\frac{J(\phi^{s+1/2}\rightarrow
\phi^{\prime})}{(1/(\sigma_{g}\sqrt{2\pi}))\exp(-(\phi^{\prime}_{v_{1},v_{2}}-\phi^{s+1/2}_{v_{1},v_{2}})^{2}/(2\sigma_{g}^{2}))}.
\end{eqnarray*}
We denote by $|B|$ the number of elements of a set $B$. All the graphs
in~$\mathcal{G}_{V}$ have the same number of neighbors, hence,
$|\operatorname{nbd}(G^{s})|=|\operatorname{nbd}(G^{\prime})|=p(p-1)/2$. Since the free
elements of $\phi^{\prime}$ are the free elements of $\phi^{s+1/2}$
and~$\phi^{\prime}_{v_{1},v_{2}}$, the Jacobian of the transformation from
$\phi^{s+1/2}$ to $\phi^{\prime}$ is equal to $1$, that is,
$J(\phi^{s+1/2}\rightarrow\phi^{\prime})=1$. Moreover, $\phi^{s+1/2}$
and $\phi^{\prime}$ have the same elements on the main diagonal and are
upper triangular, therefore, $\operatorname{det} K^{s+1/2}=\operatorname{det}
K^{\prime}$. We also have\vspace*{-3pt}
\[
\frac{J(K^{\prime} \rightarrow\phi^{\prime})}{J(K^{s+1/2}
\rightarrow\phi^{s+1/2})} = \frac{(\phi^{\prime
}_{v_{1},v_{1}})^{d_{v_{1}}^{G^{\prime}}+1}}{(\phi
^{s+1/2}_{v_{1},v_{1}})^{d_{v_{1}}^{G^{s}}+1}} = \phi^{s+1/2}_{v_{1},v_{1}}.\vspace*{-3pt}
\]
It follows that $R_{g}$ is equal to\vspace*{-3pt}
\begin{eqnarray*}
&&
\sigma_{g}\sqrt{2\pi}\phi^{s+1/2}_{v_{1},v_{1}} \frac{I_{G^{s}}(\delta,D)}{I_{G^{\prime}}(\delta,D)}
\\[-2pt]
&&\qquad{}\times
\exp\Biggl\{-\frac{1}{2}\Biggl\langle K^{\prime}-K^{s+1/2},D+\sum_{j=1}^nz^{(j)}z^{(j)T}\Biggr\rangle+ \frac{(\phi
^{\prime}_{v_{1},v_{2}}-\phi^{s+1/2}_{v_{1},v_{2}})^{2}}{2\sigma_{g}^{2}}\Biggr\}.\vspace*{-3pt}
\end{eqnarray*}

Now we examine the case when there is an edge between $v_{1}$ and
$v_{2}$ in $G^{s}$. We delete this edge from $G^{s}$ to obtain a
candidate graph $G^{\prime}$. We have $bd_{G^{\prime}}(v_{1}) =
bd_{G^{s}}(v_{1})\setminus\{ v_{2}\}$, hence,
$d_{v_{1}}^{G^{\prime}}=d_{v_{1}}^{G^{s}}-1$ and
$\nu(G^{\prime})=\nu(G^{s})\setminus\{(v_{1},v_{2})\}$. We define an
upper diagonal matrix $\phi^{\prime}$ such that
$\phi^{\prime}_{v_{1}^{\prime},v_{2}^{\prime}}=\phi
^{s+1/2}_{v_{1}^{\prime},v_{2}^{\prime}}$
for all $(v^{\prime}_{1},v^{\prime}_{2})\in\nu(G^{\prime})$. The
$(v_{1},v_{2})$ element is free in $\phi^{s+1/2}$, but it is no longer
free in $\phi^{\prime}$. The nonfree elements of $\phi^{\prime}$ are
obtained by completion with respect to the graph $G^{\prime}$. As
before, we take $K^{\prime}=(\phi^{\prime})^{T}\phi^{\prime}\in
P_{G^{\prime}}$. The dimensionality of the parameter space decreases by
$1$ as we move from $\phi^{s+1/2}$ to $\phi^{\prime}$. We obtain that
the acceptance probability of the update from $(K^{s+1/2},G^{s})$ to
$(K^{\prime},G^{\prime})$ is $\min\{R_{g}^{\prime},1\}$, where
$R_{g}^{\prime}$ is equal to\vspace*{-3pt}
\begin{eqnarray*}
&&
\bigl(\sigma_{g}\sqrt{2\pi}\phi^{s+1/2}_{v_{1},v_{1}} \bigr)^{-1} \frac{I_{G^{s}}(\delta,D)}{I_{G^{\prime}}(\delta,D)}
\\[-2pt]
&&\qquad{}\times
\exp\Biggl\{-\frac{1}{2}\Biggl\langle K^{\prime}-K^{s+1/2},D+\sum_{j=1}^nz^{(j)}z^{(j)T}\Biggr\rangle- \frac{(\phi
^{\prime}_{v_{1},v_{2}}-\phi^{s+1/2}_{v_{1},v_{2}})^{2}}{2\sigma_{g}^{2}}\Biggr\}.\vspace*{-3pt}
\end{eqnarray*}
The updated graph and the corresponding precision matrix that are
obtained at the end of this step are $G^{s+1}$ and $K^{s+1}$,
respectively.$ $

We note that our strategy for updating the precision matrix and the
graph has some similarities with the work of
\citet{giudici3green31999}. However, they focused exclusively on
decomposable graphs and perturbed elements of the covariance matrix
$K^{-1}$ that are either on its main diagonal or correspond to an edge
in the graph.

\subsection{Estimation and testing in copula Gaussian graphical models}

In high-dimensional data sets with a small number of observed samples
it is likely that the highest posterior probability graph receives only
a small (almost zero) posterior probability. Furthermore, changing a
few edges in this graph could lead to graphs with comparable posterior
probabilities. When model uncertainty is high, Bayesian model averaging
becomes key because it avoids the need to perform inference by making
an explicit choice about which edges are present or absent in the
graphs that underlie the CGGMs. This choice is not desirable since a
small sample size means lack of sufficient information. As such,
averaging over a large number of graphs is preferable even if
prediction is not the final goal.

We let $\{(G^{s},K^{s},\Upsilon^{s})\dvtx s=1,2,\ldots,S\}$ be samples from
the joint distribution~(\ref{eq:joint}), where $\Upsilon^{s}$ is the
correlation matrix corresponding with $K^{s}$---see~(\ref{eq:cork}).
These samples can be used to produce Monte Carlo estimates of functions
involving the latent variables $Z$ or the observed variables $X$. The
posterior probability that two latent variables $Z_{v_1}$ and $Z_{v_2}$
are not conditionally independent given $Z_{V\setminus\{v_1,v_2\}}$ is
the posterior inclusion probability of the edge ($v_1$,$v_2$) which is
estimated as the proportion of graphs $G^{s}$ that contain the edge
$(v_1,v_2)$.

The posterior expectation of the correlation matrix $\Upsilon$ is
estimated by the mean
$\widetilde{\Upsilon}=\frac{1}{S}\sum_{s=1}^S\Upsilon^{s}$. A zero
element of the correlation matrix $\Upsilon$ implies the independence
of $Z_{v_{1}}$ and $Z_{v_{2}}$, which in turn implies the independence
of~$X_{v_{1}}$ and $X_{v_{2}}$. We can conduct a Bayesian test of
independence of $X_{v_{1}}$ and~$X_{v_{2}}$ by considering the interval
null hypothesis $H_{0,\Upsilon}^{v_{1},v_{2}}\dvtx
|\Upsilon_{v_{1},v_{2}}|<\varepsilon$ with the alternative
$H_{1,\Upsilon}^{v_{1},v_{2}}\dvtx |\Upsilon_{v_{1},v_{2}}|\ge
\varepsilon$, where $\varepsilon>0$. Given equal apriori probabilities
of the null and alternative hypotheses, the Bayes factor
\[
B^{v_{1},v_{2}}_{\Upsilon}=p\bigl(H_{1,\Upsilon
}^{v_{1},v_{2}}|x^{(1:n)}\bigr)\bigl/p\bigl(H_{0,\Upsilon}^{v_{1},v_{2}}|x^{(1:n)}\bigr)
\]
is estimated as the number of $\Upsilon^{s}_{v_{1},v_{2}}$ whose
absolute value is above $\varepsilon$ divided by the number of
$\Upsilon^{s}_{v_{1},v_{2}}$ whose absolute value is below
$\varepsilon$.

The CDF of $X=X_V$ is estimated as
\[
\frac{1}{S}\sum_{s=1}^SC(\widehat{F}_1(x_1),\ldots,\widehat
{F}_p(x_p)|\Upsilon^{s}),
\]
where $\widehat{F}_v$ is the empirical univariate distribution of
$X_v$. If each observed variable is discrete and takes values
$\{0,1,2,\ldots\}$, their joint probability given~$\Upsilon$ is
[\citet{song22000}]
%
\begin{equation} \label{eq:copest}
\quad\quad p(X_V=x_V|\Upsilon)=\sum
_{j_1=0}^1\cdots\sum_{j_p=0}^1(-1)^{j_1+\cdots
+j_p}C(u_1^{j_1}(x_1),\ldots,u_p^{j^p}(x_p)|\Upsilon),
\end{equation}
where $u_v^{0}(x_v)=\widehat{F}_v(x_v)$ and
$u_v^{1}(x_v)=\widehat{F}_v(x_v-1)$. We define $u^1_v(0)=0$. For
example, if $X_v\in\{0,1\}$ is a binary random variable, we have
$u_v^0(1)=1$ and
$u_v^0(0)=u_v^1(1)=\frac{1}{n}\sum_{i=1}^n\delta_{\{x^{(i)}_v=0\}}$.
Here $\delta_B$ is $1$ if $B$ is true and is $0$ otherwise. Thus, the
posterior expectation of the joint probability of $X_{V}$ is estimated~%
as
\[
\widetilde{p}(X_{V}=v_{v}) = \frac{1}{S}\sum_{s=1}^{S}
p(X_V=x_V|\Upsilon^{s}).
\]

Cram\'{e}r's V [\citet{cramer41946}] is a measure of association between two
categorical variables $X_{v_{1}}$ and $X_{v_{2}}$ that take values in
the finite sets $\mathcal{I}_{v_{1}}$ and~$\mathcal{I}_{v_{2}}$,
respectively,
%
\begin{eqnarray} \label{eq:cramersv}
 \rho_{v_{1},v_{2}} &=&
\frac{1}{\min\{|\mathcal{I}_{v_{1}}|,|\mathcal{I}_{v_{2}}|\}-1}\nonumber
\\[-8pt]\\[-8pt]
&&{}\times\sum
_{x_{v_{1}}\in
\mathcal{I}_{v_{1}}}\sum_{x_{v_{2}}\in\mathcal{I}_{v_{2}}}
\frac
{p^{2}(X_{v_{1}}=x_{v_{1}},X_{v_{2}}=x_{v_{2}})}{p(X_{v_{1}}=x_{v_{1}})p(X_{v_{2}}=x_{v_{2}})}-1.\nonumber
\end{eqnarray}
Cram\'{e}r's V always takes values between $0$ and $1$, but we have
$\rho_{v_{1},v_{2}}=0$ if and only if $X_{v_{1}}$ and $X_{v_{2}}$ are
independent. The posterior expectation of~$\rho_{v_{1},v_{2}}$ is
estimated by calculating the marginal cell value
$p(X_{v_{1}}=x_{v_{1}},X_{v_{2}}=x_{v_{2}}| \Upsilon^{s})$ of
$p(X_V=x_V|\Upsilon^{s})$ for $s=1,2,\ldots,S$, calculating
$\rho^{s}_{v_{1},v_{2}}$ from (\ref{eq:cramersv}) with respect to
$p(X_{v_{1}}=x_{v_{1}},X_{v_{2}}=x_{v_{2}}| \Upsilon^{s})$ for
$s=1,2,\ldots,S$, then taking the average
$\widetilde{\rho}_{v_{1},v_{2}}=\frac{1}{S}\sum_{s=1}^{S}\rho
^{s}_{v_{1},v_{2}}$.

We can test the independence of $X_{v_{1}}$ and $X_{v_{2}}$ based on
Cram\'{e}r's V as follows. We consider the null hypothesis
$H_{0,\rho}^{v_{1},v_{2}}\dvtx \rho_{v_{1},v_{2}}<\varepsilon$ against the
alternative
$H_{1,\rho}^{v_{1},v_{2}}\dvtx \rho_{v_{1},v_{2}}\ge\varepsilon$. The
corresponding Bayes factor in favor of the alternative hypothesis is
\[
B^{v_{1},v_{2}}_{\rho}=p\bigl(H_{1,\rho
}^{v_{1},v_{2}}|x^{(1:n)}\bigr)\bigl/p\bigl(H_{0,\rho}^{v_{1},v_{2}}|x^{(1:n)}\bigr),
\]
where we assumed equal apriori probabilities of
$H_{0,\rho}^{v_{1},v_{2}}$ and $H_{1,\rho}^{v_{1},v_{2}}$. We estimate
$B^{v_{1},v_{2}}_{\rho}$ as the number of $\rho^{s}_{v_{1},v_{2}}$
above $\varepsilon$ divided by the number of $\rho^{s}_{v_{1},v_{2}}$
below $\varepsilon$. We note that \citet{dunson4xing42009} have also
used Cram\'{e}r's V to perform Bayesian testing for multivariate
categorical data in a nonparametric framework.

\begin{table}
\caption{Rochdale data from
\citet{whittaker1990}. The cells counts appear row by row in
lexicographical order with variable $h$ varying fastest and variable
$a$ varying slowest. The grand total of this table is $665$}\label{tab:rochdaledata}
\begin{tabular*}{\textwidth}{@{\extracolsep{\fill}}d{2.0}d{2.0}d{2.0}cd{2.0}d{2.0}ccd{2.0}d{2.0}ccd{2.0}d{2.0}cc@{}}\hline
$5$ & $0$ & $2$ & $1$ & $5$ & $1$ & $0$ & $0$ & $4$ & $1$ & $0$ & $0$ &
$6$ & $0$ & $2$ & $0$\\
$8$ & $0$ & $11$ & $0$ & $13$ & $0$ & $1$ & $0$ & $3$ & $0$ & $1$ & $0$
& $26$ & $0$ & $1$ & $0$\\
$5$ & $0$ & $2$ & $0$ & $0$ & $0$ & $0$ & $0$ & $0$ & $0$ & $0$ & $0$ &
$0$ & $0$ & $1$ & $0$\\
$4$ & $0$ & $8$ & $2$ & $6$ & $0$ & $1$ & $0$ & $1$ & $0$ & $1$ & $0$ &
$0$ & $0$ & $1$ & $0$\\
$17$ & $10$ & $1$ & $1$ & $16$ & $7$ & $0$ & $0$ & $0$ & $2$ & $0$ &
$0$ & $10$ & $6$ & $0$ & $0$\\
$1$ & $0$ & $2$ & $0$ & $0$ & $0$ & $0$ & $0$ & $1$ & $0$ & $0$ & $0$ &
$0$ & $0$ & $0$ & $0$\\
$4$ & $7$ & $3$ & $1$ & $1$ & $1$ & $2$ & $0$ & $1$ & $0$ & $0$ & $0$ &
$1$ & $0$ & $0$ & $0$\\
$0$ & $0$ & $3$ & $0$ & $0$ & $0$ & $0$ & $0$ & $0$ & $0$ & $0$ & $0$ &
$0$ & $0$ & $0$ & $0$\\
$18$ & $3$ & $2$ & $0$ & $23$ & $4$ & $0$ & $0$ & $22$ & $2$ & $0$ &
$0$ & $57$ & $3$ & $0$ & $0$\\
$5$ & $1$ & $0$ & $0$ & $11$ & $0$ & $1$ & $0$ & $11$ & $0$ & $0$ & $0$
& $29$ & $2$ & $1$ & $1$\\
$3$ & $0$ & $0$ & $0$ & $4$ & $0$ & $0$ & $0$ & $1$ & $0$ & $0$ & $0$ &
$0$ & $0$ & $0$ & $0$\\
$1$ & $1$ & $0$ & $0$ & $0$ & $0$ & $0$ & $0$ & $0$ & $0$ & $0$ & $0$ &
$0$ & $0$ & $0$ & $0$\\
$41$ & $25$ & $0$ & $1$ & $37$ & $26$ & $0$ & $0$ & $15$ & $10$ & $0$ &
$0$ & $43$ & $22$ & $0$ & $0$\\
$0$ & $0$ & $0$ & $0$ & $2$ & $0$ & $0$ & $0$ & $0$ & $0$ & $0$ & $0$ &
$3$ & $0$ & $0$ & $0$\\
$2$ & $4$ & $0$ & $0$ & $2$ & $1$ & $0$ & $0$ & $0$ & $1$ & $0$ & $0$ &
$2$ & $1$ & $0$ & $0$\\
$0$ & $0$ & $0$ & $0$ & $0$ & $0$ & $0$ & $0$ & $0$ & $0$ & $0$ & $0$ &
$0$ & $0$ & $0$ & $0$\\
\hline
\end{tabular*}
\end{table}

In the two examples discussed in Section \ref{sec:examples} we chose to
test independence of each pair of variables based on Cram\'{e}r's V since
this measure takes into account the univariate distributions of the
observed variables.

\section{Examples} \label{sec:examples}

In this section we apply copula GGMs to analyze two multivariate data
sets with high relevance in the social science literature. In the
supplementary material [\citet{dobrabenkoskibaoassuppv2010}] we provide
C++ code and the data sets that are needed to replicate the numerical
results that follow.

\subsection{The Rochdale data}

We consider a social survey data set previously analyzed in
\citet{whittaker1990}---see Table \ref{tab:rochdaledata}. This
observational study was conducted in Rochdale and attempted to assess
the relationships among factors affecting women's economic activity.
The eight variables are as follows: $a$, wife economically active
(no, yes); $b$, age of wife $>38$ (no, yes); $c$,~husband unemployed
(no, yes); $d$, child $\le4$ (no, yes); $e$, wife's education,
high-school$+$ (no, yes); $f$, husband's education, high-school$+$ (no, yes);
$g$,~Asian origin (no, yes); $h$, other household member working
(no, yes). The resulting $2^8$ cross-classification has~$165$ counts of
zero, while $217$ cells contain small positive counts smaller than~$3$.
There are quite a few counts larger than $30$ or even $50$.

Since the sample size is only $665$, this table is sparse.
\citet{whittaker1990} argues that higher-order interactions involving
more than two variables should not be included in any log-linear model
that is fit to this data set. He subsequently studies two log-linear
models: the all two-way interaction model whose minimal sufficient
statistics are all the $28$ two-way marginals and the model whose
minimal sufficient statistics are the two-way marginals corresponding
with the pairs of variables
%
\begin{equation}
\label{eq:loglinearwhittaker}
\{ fg,ef,dh,dg,cg,cf,ce,bh,be,bd,ag,ae,ad,ac\}.
\end{equation}

\begin{figure}

\includegraphics{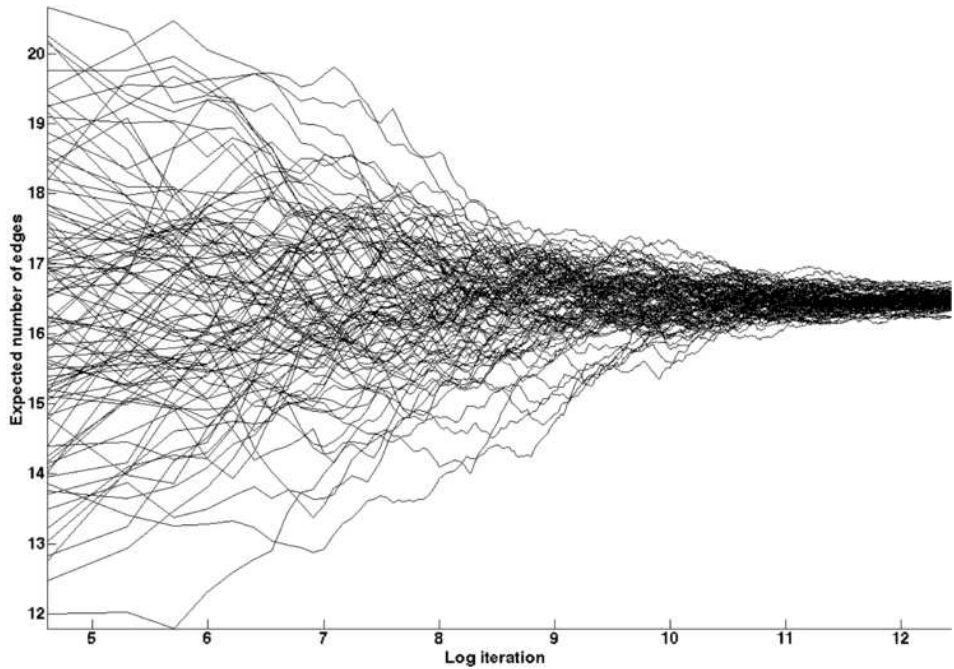}

\caption{Estimates of the posterior expected
number of edges in the CGGMs for the Rochdale data.}\label{fig:chainrochdale}
\end{figure}

We ran the Markov chain Monte Carlo sampler from Section~\ref{sec:mcmc}
for 250,000 iterations from 100 random starting graphs. The burn-in time
was 25,000 iterations. Convergence to the stationary distribution
(\ref{eq:joint}) is illustrated in Figure~\ref{fig:chainrochdale} that
gives the posterior expected number of edges in the CGGM graphs across
iterations for each chain. The sampled graphs have on average $16.5$
edges which represent approximately $59\%$ of the total number of
possible edges. By comparison, the log-linear model
(\ref{eq:loglinearwhittaker}) has $14$ minimal sufficient statistics.

In order to show the importance of modeling the conditional
independence relationships among the latent variables using graphs, we
have also employed the copula estimation approach proposed by
\citet{hoff22007}---see equation (\ref{eq:hoff}). Hoff's method is
equivalent to starting the Markov chain from Section~\ref{sec:mcmc} at
the full graph and never updating this graph by skipping step 3 of the
algorithm. Moreover, updating the precision matrix from step 2 is
performed by direct sampling from the Wishart posterior $W_{p}(
\delta+n,D+\sum_{j=1}^{n}z^{(j)}z^{(j)T})$. This simplified Markov
chain was run for $25$ million iterations and henceforth is called the
Copula-Full model.

We compare the expected cell counts of the all two-way interaction
log-linear model, the log-linear model (\ref{eq:loglinearwhittaker}),
the Copula-Full model and the CGGMs. Table~\ref{tab:rochdale} shows the
cells containing the $20$ largest observed counts together with their
corresponding estimates. It is remarkable that the CGGMs perform as
well as the all two-way interaction model for the largest cell count
$57$. The squared errors between the observed counts and the expected
cell counts for all the $256$ cells in the table are the following:
$284.79$ for the all two-way interaction model, $407.04$ for the CGGMs,
$905.78$ for the model~(\ref{eq:loglinearwhittaker}) and $1919.15$ for
the Copula-Full model.

\begin{table}
\caption{Expected cell counts for the top $20$
largest counts cells associated with the all two-way interaction
log-linear model, Whittaker's log-linear model (\protect\ref{eq:loglinearwhittaker}), the Copula-Full model and the CGGMs in the
Rochdale data. Here 1 stands for no and 2 stands for yes}\label{tab:rochdale}
\vspace*{-3pt}
\begin{tabular*}{\textwidth}{@{\extracolsep{\fill}}lcd{2.2}d{2.2}d{2.2}d{2.2}@{}}
\hline
\textbf{Cell} & \textbf{Observed} & \multicolumn{1}{c}{\textbf{All two-way}} & \multicolumn{1}{c}{\textbf{Whittaker}} & \multicolumn{1}{c}{\textbf{Copula-Full}}&\multicolumn{1}{c@{}}{\textbf{CGGMs}}\\
\hline
2 1 1 1 2 2 1 1 & 57 & 56.78 & 52.08 & 39.43 & 56.80\\
2 2 1 1 2 2 1 1 & 43 & 44.61 & 40.97 & 36.58 & 47.55\\
2 2 1 1 1 1 1 1 & 41 & 36.40 & 36.32 & 30.48 & 36.12\\
2 2 1 1 1 2 1 1 & 37 & 38.77 & 36.92 & 35.33 & 36.61\\
2 1 1 2 2 2 1 1 & 29 & 33.29 & 39.06 & 17.85 & 32.40\\
1 1 1 2 2 2 1 1 & 26 & 20.36 & 9.63 & 9.53 & 18.03\\
2 2 1 1 1 2 1 2 & 26 & 23.68 & 22.89 & 15.67 & 24.54\\
2 2 1 1 1 1 1 2 & 25 & 28.12 & 22.52 & 15.11 & 27.63\\
2 1 1 1 1 2 1 1 & 23 & 22.73 & 20.06 & 26.51 & 22.76\\
2 1 1 1 2 1 1 1 & 22 & 19.22 & 16.54 & 17.15 & 16.75\\
2 2 1 1 2 2 1 2 & 22 & 22.85 & 25.41 & 13.96 & 24.63\\
2 1 1 1 1 1 1 1 & 18 & 21.54 & 19.74 & 21.02 & 20.85\\
1 2 1 1 1 1 1 1 & 17 & 15.06 & 16.02 & 15.13 & 15.71\\
1 2 1 1 1 2 1 1 & 16 & 14.65 & 16.28 & 14.3 &12.18\\
2 2 1 1 2 1 1 1 & 15 & 14.96 & 13.01 & 17.36 & 15.07\\
1 1 1 2 1 2 1 1 & 13 & 12.06 & 6.63 & 8.46 & 10.92\\
2 1 1 2 2 1 1 1 & 11 & 7.70 & 12.40 & 7.36 & 8.52\\
2 1 1 2 1 2 1 1 & 11 & 10.50 & 15.05 & 11 & 10.48\\
1 1 1 2 1 1 2 1 & 11 & 8.08 & 6.72 & 1.53 & 6.31\\
\hline
\end{tabular*}
\vspace*{-3pt}
\end{table}

\begin{table}[b]
\vspace*{-3pt}
\caption{Estimated correlations
(elements under the main diagonal) and posterior inclusion
probabilities of edges (elements above the main diagonal) associated
with the CGGMs in the Rochdale data}\label{tab:rochdale-edges-corr}
\vspace*{-3pt}
\begin{tabular*}{\textwidth}{@{\extracolsep{\fill}}ld{2.2}d{2.2}d{2.2}d{2.2}d{2.2}d{2.2}d{2.2}c@{}}
\hline
& \multicolumn{1}{c}{$\bolds{a}$} & \multicolumn{1}{c}{$\bolds{b}$} & \multicolumn{1}{c}{$\bolds{c}$} & \multicolumn{1}{c}{$\bolds{d}$} & \multicolumn{1}{c}{$\bolds{e}$} & \multicolumn{1}{c}{$\bolds{f}$} & \multicolumn{1}{c}{$\bolds{g}$} & $\bolds{h}$\\
\hline
\textit{a} & \multicolumn{1}{c}{---} & 0.93 & 0.67 & 0.92 & 0.32 & 0.42 & 1 & 0.26\\
\textit{b} & 0.15 & \multicolumn{1}{c}{---} & 0.27 & 1 & 0.88 & 0.29 & 0.70 & 0.96\\
\textit{c} & -0.52 & -0.02 &\multicolumn{1}{c}{---} & 0.29 & 0.91 & 0.35 & 0.85 & 0.25\\
\textit{d} & -0.46 & -0.79 & 0.19 & \multicolumn{1}{c}{---} & 0.37 & 0.59 & 0.66 & 0.50 \\
\textit{e} & 0.30 & -0.28 & -0.48 & 0.12 & \multicolumn{1}{c}{---} & 0.98 & 0.58 & 0.17\\
\textit{f} & 0.22 & -0.11 & -0.35 & 0.04 & 0.46 & \multicolumn{1}{c}{---} & 0.82 & 0.22\\
\textit{g} & -0.71 & -0.31 & 0.57 & 0.51 & -0.34 & -0.37 & \multicolumn{1}{c}{---} & 0.32 \\
\textit{h} & 0.12 & 0.63 & 0.01 & -0.54 & -0.19 & -0.10 & -0.18 & \multicolumn{1}{c@{}}{---} \\
\hline
\end{tabular*}
\end{table}

In Table \ref{tab:rochdale-edges-corr} we show the pairwise
correlations $\Upsilon_{v_{1},v_{2}}$ and the posterior inclusion
probabilities of edges $(v_{1},v_{2})$ for any two latent variables
$Z_{v_{1}}$ and~$Z_{v_{2}}$ as estimated using the CGGMs. In Table
\ref{tab:rochdale-edges-corr-full} we give the estimates of the
pairwise correlations $\Upsilon_{v_{1},v_{2}}$ obtained using the
Copula-Full model. We see that the absolute values of these estimates
are significantly smaller than corresponding absolute values of the
CGGMs estimates. We show the dependence structure of the observed
variables in Tables \ref{tab:rochdale-cramerv} and
\ref{tab:rochdale-cramerv-full}. We give the posterior means of
Cram\'{e}r's V $\rho_{v_{1},v_{2}}$ and estimates of the posterior
probabilities $p(H_{1,\rho}^{v_{1},v_{2}}|x^{(1:n)})$ with
$H_{1,\rho}^{v_{1},v_{2}}\dvtx \rho_{v_{1},v_{2}}>0.1$. By contrasting the
estimates obtained using CGGMs and the Copula-Full model, we clearly
see that conditioning on the full graph is quite disadvantageous: the
Cram\'{e}r's~V associations are severely underestimated and, subsequently,
all the posterior probabilities $p(H_{1,\rho}^{v_{1},v_{2}}|x^{(1:n)})$
are almost zero under the full graph. The CGGMs take every possible
graph into account and the corresponding estimates are produced by
Bayesian model averaging across all graphs. This leads to more
appropriate results as evidenced in Tables
\ref{tab:rochdale-edges-corr}--\ref{tab:rochdale-cramerv-full}.

\begin{table}
\caption{Estimated correlations
(elements under the main diagonal) associated with the Copula-Full
model in the Rochdale data}\label{tab:rochdale-edges-corr-full}
\begin{tabular*}{\textwidth}{@{\extracolsep{\fill}}ld{2.2}d{2.2}d{2.2}d{2.2}d{2.2}d{2.2}d{2.2}c@{}}
\hline
& \multicolumn{1}{c}{$\bolds{a}$} & \multicolumn{1}{c}{$\bolds{b}$} & \multicolumn{1}{c}{$\bolds{c}$} & \multicolumn{1}{c}{$\bolds{d}$} & \multicolumn{1}{c}{$\bolds{e}$} & \multicolumn{1}{c}{$\bolds{f}$} & \multicolumn{1}{c}{$\bolds{g}$} & $\bolds{h}$\\
\hline
\textit{a} & \multicolumn{1}{c}{---} & \\
\textit{b} & 0.08 & \multicolumn{1}{c}{---} & \\
\textit{c} & -0.17 & -0.02 & \multicolumn{1}{c}{---} & \\
\textit{d} & -0.20 & -0.35 & 0.06 & \multicolumn{1}{c}{---} & \\
\textit{e} & 0.15 & -0.15 & -0.15 & 0.05 & \multicolumn{1}{c}{---} & \\
\textit{f} & 0.10 & -0.06 & -0.13 & 0.02 & 0.24 & \multicolumn{1}{c}{---} & \\
\textit{g} & -0.18 & -0.08 & 0.13 & 0.13 & -0.09 & -0.11 & \multicolumn{1}{c}{---} & \\
\textit{h} & 0.05 & 0.27 & 0.01 & -0.18 & -0.08 & -0.06 & -0.04 & --- \\
\hline
\end{tabular*}
\end{table}

\begin{table}[b]
\caption{Estimated Cram\'{e}r's V
associations (elements under the main diagonal) and posterior
probabilities $p(H_{1,\rho}|x^{(1:n)})$ (elements above the main
diagonal) associated with the CGGMs in the Rochdale data}\label{tab:rochdale-cramerv}
\begin{tabular*}{\textwidth}{@{\extracolsep{\fill}}ld{1.2}d{1.2}d{1.2}d{1.2}d{1.2}d{1.2}d{1.2}d{1.2}@{}}
\hline
& \multicolumn{1}{c}{$\bolds{a}$} & \multicolumn{1}{c}{$\bolds{b}$} & \multicolumn{1}{c}{$\bolds{c}$} & \multicolumn{1}{c}{$\bolds{d}$} & \multicolumn{1}{c}{$\bolds{e}$} & \multicolumn{1}{c}{$\bolds{f}$} & \multicolumn{1}{c}{$\bolds{g}$} & \multicolumn{1}{c@{}}{$\bolds{h}$}\\
\hline
\textit{a} & \multicolumn{1}{c}{---} & 0 & 0.19 & 0.22 & 0 & 0 & 0.83 & 0\\
\textit{b} & 0.01 & \multicolumn{1}{c}{---} & 0 & 1 & 0 & 0 & 0 & 0.94\\
\textit{c} & 0.08 & 0 & \multicolumn{1}{c}{---} & 0 & 0 & 0 & 0.42 & 0\\
\textit{d} & 0.08 & 0.24 & 0.01 & \multicolumn{1}{c}{---} & 0 & 0 & 0.07 & 0 \\
\textit{e} & 0.04 & 0.03 & 0.05 & 0.01 & \multicolumn{1}{c}{---} & 0.35 & 0 & 0\\
\textit{f} & 0.02 & 0.01 & 0.03 & 0 & 0.09 & \multicolumn{1}{c}{---} & 0 & 0\\
\textit{g} & 0.12 & 0.02 & 0.09 & 0.07 & 0.02 & 0.03 & \multicolumn{1}{c}{---} & 0 \\
\textit{h} & 0.01 & 0.14 & 0 & 0.06 & 0.01 & 0 & 0 & \multicolumn{1}{c@{}}{---} \\
\hline
\end{tabular*}
\end{table}

\citet{whittaker1990}, page 282, argues that the strongest pairwise
interaction in the Rochdale data is $(b,d)$, followed by $(b,h)$,
$(e,f)$ and $(a,g)$. In Table~\ref{tab:rochdale-edges-corr} we see that
the top four posterior inclusion probabilities in the CGGMs are as
follows: $1$ for $(b,d)$, $0.96$ for $(b,h)$, $0.98$ for $(e,f)$ and
$1$ for $(a,g)$. The strongest associations in the observed variables
space as measured by Cram\'{e}r's V are the following: $(b,d)$, $(b,h)$,
$(a,g)$, $(e,f)$ and $(c,g)$. The interaction between $c$ and $g$ is
also present in the log-linear model (\ref{eq:loglinearwhittaker}).

\begin{table}
\caption{Estimated Cram\'{e}r's V
associations (elements under the main diagonal) and posterior
probabilities $p(H_{1,\rho}|x^{(1:n)})$ (elements above the main
diagonal) associated with the Copula-Full model in the Rochdale data}\label{tab:rochdale-cramerv-full}
\begin{tabular*}{\textwidth}{@{\extracolsep{\fill}}ld{1.2}d{1.2}d{1.2}cd{1.2}ccc@{}}
\hline
& \multicolumn{1}{c}{$\bolds{a}$} & \multicolumn{1}{c}{$\bolds{b}$} & \multicolumn{1}{c}{$\bolds{c}$} & \multicolumn{1}{c}{$\bolds{d}$} & \multicolumn{1}{c}{$\bolds{e}$} & \multicolumn{1}{c}{$\bolds{f}$} & \multicolumn{1}{c}{$\bolds{g}$} & \multicolumn{1}{c@{}}{$\bolds{h}$}\\
\hline
\textit{a} & \multicolumn{1}{c}{---} & 0 & 0 & 0 & 0 & 0 & 0 & 0\\
\textit{b} & 0 & \multicolumn{1}{c}{---} & 0 & 0 & 0 & 0 & 0 & 0\\
\textit{c} & 0.01 & 0 & \multicolumn{1}{c}{---} & 0 & 0 & 0 & 0 & 0\\
\textit{d} & 0.02 & 0.04 & 0 & --- & 0 & 0 & 0 & 0 \\
\textit{e} & 0.01 & 0.01 & 0.01 & 0 & \multicolumn{1}{c}{---} & 0 & 0 & 0\\
\textit{f} & 0.01 & 0 & 0.01 & 0 & 0.03 & --- & 0 & 0\\
\textit{g} & 0.01 & 0 & 0 & 0 & 0 & 0 & --- & 0 \\
\textit{h} & 0 & 0.03 & 0 & 0 & 0 & 0 & 0 & --- \\
\hline
\end{tabular*}
\end{table}

Of particular interest is the determination of the factors that
influence variable $a$---the wife's economic activity. From Table
\ref{tab:rochdale-cramerv} we see that variables~$c$, $d$ and $g$ are
the only variables with a strictly positive posterior probability that
their Cram\'{e}r's V association with variable $a$ is greater than~$0.1$.
The largest Cram\'{e}r's V association is $\widetilde{\rho}_{a,g}=0.12$,
followed by $\widetilde{\rho}_{a,c}=0.08$ and
$\widetilde{\rho}_{a,d}=0.08$. The corresponding estimated correlations
from Table \ref{tab:rochdale-edges-corr} show a~negative relationship
between $a$ and each of these three variables. \citet{whittaker1990}
determines which variables influence $a$ by considering the log-linear
model $ac|ad|ae|ag$ induced by the generators of model~%
(\ref{eq:loglinearwhittaker}) that involve $a$. Using maximum
likelihood estimation of log-linear parameters, Whittaker obtains the
following estimates of the logistic regression of $a$ on $c$, $d$, $e$
and~$g$:
%
\begin{equation}\label{eq:logisticwhit}
\log\frac{p(a=1|c,d,e,g)}{p(a=0|c,d,e,g)}  =  \mathrm{const.} - 1.33
c - 1.32 d + 0.69 e - 2.17 g.
\end{equation}
Equation (\ref{eq:logisticwhit}) seems to support our findings based on
CGGMs, as it indicates a negative association between $(a,c)$, $(a,d)$,
$(a,g)$, and a positive association between $(a,e)$. Moreover, the
association between $a$ and $e$ is the weakest of the four. The CGGMs
estimate $\widetilde{\rho}_{a,e}=0.04$ which is about half of~%
$\widetilde{\rho}_{a,c}$ or $\widetilde{\rho}_{a,d}$. The absolute
values of the regression coefficients in (\ref{eq:logisticwhit}) share
the same pattern.

We remark that Table \ref{tab:rochdale-edges-corr} reports a posterior
inclusion probability equal to $0.93$ for the edge $(a,b)$. However,
the CGGMs estimate the pairwise correlation $\Psi_{a,b}$ to be $0.15$
and the Cram\'{e}r's V association $\rho_{a,b}$ to be $0.01$. Therefore,
the CGGMs do not seem to indicate a relevant interaction between
variables $a$ and $b$ which is in line with Whittaker's findings who
did not include an interaction term $ab$ in model
(\ref{eq:loglinearwhittaker}). This represents an example where an edge
vanishes as we move from the latent variables space to the observed
variables space. We would expect the opposite to happen in most
applications, that is, edges or associations could be lost when moving
from the observed to the latent variables.

\begin{table}[b]
\caption{Expected cell counts for the top six largest
counts cells in the NLTCS data. We report the results obtained from the
GoM model [\citet{erosheva2008}], the LC model [\citet{fien1hers1rina1zhou12008}] and the CGGMs. Here 1 stands for healthy
and 2 stands for disabled}\label{tab:nltcs}
\begin{tabular*}{\textwidth}{@{\extracolsep{\fill}}ld{4.0}d{4.0}d{4.2}d{4.2}@{}}
\hline
\textbf{Cell} & \multicolumn{1}{c}{\textbf{Observed}} & \multicolumn{1}{c}{\textbf{GoM}} & \multicolumn{1}{c}{\textbf{LC}} & \multicolumn{1}{c@{}}{\textbf{CGGMs}}\\
\hline
1 1 1 1 1 1 1 1 1 1 1 1 1 1 1 1 & 3853 & 3269 & 3836.01 & 3767.76\\
1 1 1 1 1 2 1 1 1 1 1 1 1 1 1 1 & 1107 & 1010 & 1111.51 & 1145.86\\
2 2 2 2 2 2 2 2 2 2 2 2 2 2 2 2 & 660 & 612 & 646.39 & 574.76\\
1 1 1 1 1 1 1 1 1 1 1 2 1 1 1 1 & 351 & 331 & 360.52 & 452.75\\
1 1 1 1 1 1 2 1 1 1 1 2 1 1 1 1 & 303 & 273 & 285.27 & 350.24\\
1 1 1 1 2 1 1 1 1 1 1 1 1 1 1 1 & 216 & 202 & 220.47 & 202.12\\
\hline
\end{tabular*}
\end{table}

\subsection{The NLTCS functional disability data} \label{section:nltcsdata}

We come back to the $2^{16}$ functional disability table introduced in
Section \ref{sec:intro}. \citet{dobrfrosdffiendf2003} analyze these
data from a disclosure limitation perspective, while
\citet{fien1hers1rina1zhou12008} develop latent class (LC) models that
are very similar to the Grade of Membership (GoM) models of
\citet{erosheva2008}. The need to consider alternatives to log-linear
models for the NLTCS data comes from the severe imbalance that exists
among the cell counts in this table. The largest cell count is $3853$,
but most of the cells (62,384 or $95.19\%$) contain counts of zero,
while $1729$ ($2.64\%$) contain counts of $1$ and $1499$ ($0.76\%$)
contain counts of $2$. There are 24 cells with counts larger than 100,
which accounts for 42\% of the observed sample size 21,574. This gives
a very small mean number of observations per cell of $0.33$, which is
indicative of an extremely high degree of sparsity that is
characteristic of high-dimensional categorical data.

\begin{figure}

\includegraphics{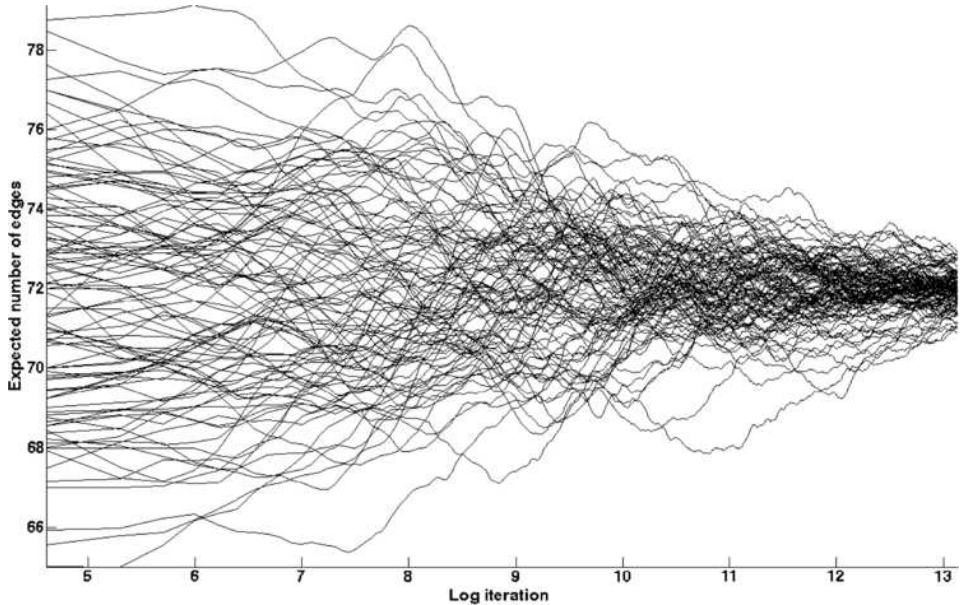}

\caption{Estimates of the posterior expected number of edges in the
CGGMs for the NLTCS functional disability data.}\label{fig:chainnltcs}
\end{figure}

We ran 100 replicates of the Markov chain Monte Carlo sampler from
Section~\ref{sec:mcmc} for 500,000 iterations with a burn-in time of
50,000 iterations. Figure~\ref{fig:chainnltcs} shows the convergence of
these Markov chains to the joint distribution (\ref{eq:joint}). The
mean number of edges of the sampled graphs is $72$ or $60\%$ of the
total number of edges. Table \ref{tab:nltcs} compares the expected cell
values of the six largest counts as estimated with the Grade of
Membership (GoM) model of \citet{erosheva2008}, the latent class (LC)
model of \citet{fien1hers1rina1zhou12008} and the \mbox{CGGMs}. All three
models seem to perform comparably well in terms of capturing the
underlying dependency patterns that lead to the largest counts in this
$2^{16}$ table.

In Table \ref{tab:nltcs-edges-corr} we show the association structure
of the latent variables $Z$. We give posterior estimates of the
pairwise correlations $\Upsilon_{v_{1},v_{2}}$ and posterior inclusion
probabilities for each edge $(v_{1},v_{2})$. All the estimates of the
pairwise correlations are quite large and strictly positive, which is
intuitively correct: the ability to perform any activity of daily
living is positively correlated with the ability to perform any other
activity. In Table \ref{tab:nltcs-cramerv} we show the association
structure of the observed variables $X$. For every pair $X_{v_{1}}$ and
$X_{v_{2}}$, we give the posterior means of $\rho_{v_{1},v_{2}}$ and
estimates of the posterior probabilities
$p(H_{1,\rho}^{v_{1},v_{2}}|x^{(1:n)})$ with
$H_{1,\rho}^{v_{1},v_{2}}\dvtx \rho_{v_{1},v_{2}}>0.1$. The Cram\'{e}r's~V
values indicate that independence is unlikely to hold for any pair of
observed variables, which is consistent with the large positive
correlations we estimated in the latent space. In fact, $88$ pairs of
observed variables have a~Bayes factor $B^{v_{1},v_{2}}_{\rho}$ greater
than $100$, which constitutes strong evidence in favor of the
hypothesis $H_{1,\rho}^{v_{1},v_{2}}$ [\citet{kass2raftery21995}]. Thus,
the NLTCS data shows that approximately $73\%$ pairs of ADLs and IADLs
are certainly not independent of each other.

\begin{sidewaystable}
\tablewidth=550pt
\caption{Estimated correlations
(elements under the main diagonal) and posterior inclusion
probabilities of edges (elements above the main diagonal) in the NLTCS data}\label{tab:nltcs-edges-corr}
\begin{tabular*}{\textwidth}{@{\extracolsep{\fill}}ld{1.2}d{1.2}d{1.2}d{1.2}d{1.2}d{1.2}d{1.2}d{1.2}d{1.2}d{1.2}d{1.2}d{1.2}d{1.2}d{1.2}d{1.2}d{1.2}@{}}
\hline
& \multicolumn{6}{c}{\textbf{ADL}} & \multicolumn{10}{c@{}}{\textbf{IADL}}\\[-5pt]
& \multicolumn{6}{c}{\hrulefill} & \multicolumn{10}{c@{}}{\hrulefill}\\
& \multicolumn{1}{c}{\textbf{1}} & \multicolumn{1}{c}{\textbf{2}} & \multicolumn{1}{c}{\textbf{3}} & \multicolumn{1}{c}{\textbf{4}} & \multicolumn{1}{c}{\textbf{5}} & \multicolumn{1}{c}{\textbf{6}} & \multicolumn{1}{c}{\textbf{1}} & \multicolumn{1}{c}{\textbf{2}} & \multicolumn{1}{c}{\textbf{3}} & \multicolumn{1}{c}{\textbf{4}} & \multicolumn{1}{c}{\textbf{5}} & \multicolumn{1}{c}{\textbf{6}} & \multicolumn{1}{c}{\textbf{7}} & \multicolumn{1}{c}{\textbf{8}} & \multicolumn{1}{c}{\textbf{9}} & \multicolumn{1}{c@{}}{\textbf{10}} \\ \hline
ADL&&&&&&&&&&&&&&&&\\
\quad \phantom{0}1 & \multicolumn{1}{c}{---} & 1 & 1 & 0.24 & 0.46 & 0.42 & 1 &0.68 & 0.87 & 0.98 & 0.33 & 1 & 0.23 & 0.46 & 0.15 & 1\\
\quad \phantom{0}2 & 0.72 & \multicolumn{1}{c}{---} & 1 & 0.19 & 0.42 & 0.94 & 1 & 0.10 & 0.18 & 0.09 &0.10 & 0.76 & 1 & 0.23 & 0.21 & 0.17\\
\quad \phantom{0}3 & 0.78 & 0.74 & \multicolumn{1}{c}{---} & 1 & 0.36 & 1 & 1 & 0.13 & 0.50 & 0.77 &0.10 & 0.74 & 0.13 & 0.78 & 0.24 & 0.16\\
\quad \phantom{0}4 & 0.51 & 0.54 & 0.64 & \multicolumn{1}{c}{---} & 1 & 1 & 0.28 & 0.16 & 1 & 1 & 0.12 &0.20 & 0.36 & 1 & 0.14 & 0.77\\
\quad \phantom{0}5 & 0.33 & 0.43 & 0.41 & 0.66 & \multicolumn{1}{c}{---} & 0.44 & 0.15 & 0.54 & 0.30 &0.95 & 0.18 & 1 & 0.81 & 1 & 0.96 & 1\\
\quad \phantom{0}6 & 0.62 & 0.65 & 0.73 & 0.82 & 0.66 & \multicolumn{1}{c}{---} & 1 & 1 & 0.34 & 0.82 &0.10 & 0.63 & 0.93 & 0.81 & 0.16 & 0.27\\
IADL&&&&&&&&&&&&&&&&\\
\quad \phantom{0}1 & 0.74 & 0.77 & 0.76 & 0.68 & 0.58 & 0.83 & \multicolumn{1}{c}{---} & 1 & 1 & 0.67 & 0.19 & 0.20 & 0.19 & 0.21 & 0.95 & 1\\
\quad \phantom{0}2 & 0.64 & 0.69 & 0.68 & 0.68 & 0.62 & 0.82 & 0.88 & \multicolumn{1}{c}{---} & 1 & 1 &0.30 & 0.13 & 0.19 & 0.27 & 0.55 & 0.72\\
\quad \phantom{0}3 & 0.65 & 0.71 & 0.66 & 0.62 & 0.61 & 0.79 & 0.90 & 0.90 & \multicolumn{1}{c}{---} & 1& 0.16 & 0.23 & 1 & 0.31 & 0.92 & 0.27\\
\quad \phantom{0}4 & 0.49 & 0.58 & 0.55 & 0.66 & 0.64 & 0.76 & 0.78 & 0.83 & 0.87 & \multicolumn{1}{c}{---} & 0.12 & 1 & 0.42 & 0.74 & 0.23 & 0.44\\
\quad \phantom{0}5 & 0.45 & 0.56 & 0.48 & 0.52 & 0.65 & 0.60 & 0.65 & 0.63 & 0.67 &0.61 & \multicolumn{1}{c}{---} & 1 & 1 & 1 & 0.97 & 0.65\\
\quad \phantom{0}6 & 0.45 & 0.59 & 0.52 & 0.56 & 0.64 & 0.64 & 0.68 & 0.66 & 0.70 &0.66 & 0.79 & \multicolumn{1}{c}{---} & 1 & 0.16 & 0.11 & 1\\
\quad \phantom{0}7 & 0.60 & 0.70 & 0.60 & 0.54 & 0.57 & 0.65 & 0.76 & 0.71 & 0.77 &0.66 & 0.79 & 0.79 & \multicolumn{1}{c}{---} & 0.33 & 1 & 1\\
\quad \phantom{0}8 & 0.39 & 0.50 & 0.43 & 0.56 & 0.87 & 0.63 & 0.62 & 0.64 & 0.66 &0.64 & 0.77 & 0.72 & 0.71 & \multicolumn{1}{c}{---} & 1 & 0.84\\
\quad \phantom{0}9 & 0.48 & 0.57 & 0.49 & 0.55 & 0.74 & 0.64 & 0.67 & 0.68 & 0.71 &0.65 & 0.79 & 0.75 & 0.80 & 0.89 & \multicolumn{1}{c}{---} & 1\\
\quad 10 & 0.65 & 0.69 & 0.63 & 0.54 & 0.52 & 0.65 & 0.77 & 0.70 & 0.75 &0.63 & 0.74 &0.75 & 0.87 & 0.68 & 0.77 & \multicolumn{1}{c}{---} \\
\hline
\end{tabular*}
\end{sidewaystable}

\begin{sidewaystable}
\tablewidth=550pt
\caption{Estimated Cram\'{e}r's V associations
(elements under the main diagonal) and posterior probabilities
$p(H_{1,\rho}|x^{(1:n)})$ (elements above the main diagonal) in the
NLTCS data}\label{tab:nltcs-cramerv}
\begin{tabular*}{\textwidth}{@{\extracolsep{\fill}}ld{1.2}d{1.2}d{1.2}d{1.2}d{1.2}d{1.2}d{1.2}d{1.2}d{1.2}d{1.2}d{1.2}d{1.2}d{1.2}d{1.2}d{1.2}d{1.2}@{}}
\hline
& \multicolumn{6}{c}{\textbf{ADL}} & \multicolumn{10}{c@{}}{\textbf{IADL}}\\[-5pt]
& \multicolumn{6}{c}{\hrulefill} & \multicolumn{10}{c@{}}{\hrulefill}\\
& \multicolumn{1}{c}{\textbf{1}} & \multicolumn{1}{c}{\textbf{2}} & \multicolumn{1}{c}{\textbf{3}} & \multicolumn{1}{c}{\textbf{4}} & \multicolumn{1}{c}{\textbf{5}} & \multicolumn{1}{c}{\textbf{6}} & \multicolumn{1}{c}{\textbf{1}} & \multicolumn{1}{c}{\textbf{2}} & \multicolumn{1}{c}{\textbf{3}} & \multicolumn{1}{c}{\textbf{4}} & \multicolumn{1}{c}{\textbf{5}} & \multicolumn{1}{c}{\textbf{6}} & \multicolumn{1}{c}{\textbf{7}} & \multicolumn{1}{c}{\textbf{8}} & \multicolumn{1}{c}{\textbf{9}} & \multicolumn{1}{c@{}}{\textbf{10}} \\ \hline
ADL&&&&&&&&&&&&&&&&\\
\quad \phantom{0}1 & \multicolumn{1}{c}{---} & 1 & 1 & 0 & 0 & 0.61 & 1 & 1 & 1 &0 & 0 & 0 & 0.99 &0 & 0 & 1\\
\quad \phantom{0}2 & 0.21 & \multicolumn{1}{c}{---} & 1 & 0.43 & 0 & 1 & 1 & 1 & 1 & 0 & 0.99 & 1 & 1 &0.08 & 1 & 1\\
\quad \phantom{0}3 & 0.26 & 0.25 & \multicolumn{1}{c}{---} & 1 & 0 & 1 & 1 & 1 & 1 & 0 & 0.05 & 0.45 & 1& 0 & 0.14 & 0.99\\
\quad \phantom{0}4 & 0.07 & 0.10 & 0.15 & \multicolumn{1}{c}{---} & 1 & 1 & 1 & 1 & 1 & 1 & 0.38 & 1 &0.32 & 1 & 0.98 & 0\\
\quad \phantom{0}5 & 0.03 & 0.06 & 0.05 & 0.21 & \multicolumn{1}{c}{---} & 1 & 0.99 & 1 & 0.98 & 1 & 1 &1 & 0.32 & 1 & 1 & 0\\
\quad \phantom{0}6 & 0.10 & 0.14 & 0.20 & 0.38 & 0.21 & \multicolumn{1}{c}{---} & 1 & 1 & 1 & 1 & 1 & 1& 1 & 1 & 1 & 0.03\\
IADL&&&&&&&&&&&&&&&&\\
\quad \phantom{0}1 & 0.21 & 0.28 & 0.28 & 0.18 & 0.11 & 0.28 & \multicolumn{1}{c}{---} & 1 & 1 & 1 & 1 & 1 & 1 & 1 & 1 & 1\\
\quad \phantom{0}2 & 0.14 & 0.19 & 0.19 & 0.21 & 0.16 & 0.34 & 0.43 & \multicolumn{1}{c}{---} & 1 & 1 &1 & 1 & 1 & 1 & 1 & 1\\
\quad \phantom{0}3 & 0.16 & 0.22 & 0.18 & 0.13 & 0.11 & 0.22 & 0.48 & 0.40 & \multicolumn{1}{c}{---} & 1& 1 & 1 & 1 & 1 & 1 & 1\\
\quad \phantom{0}4 & 0.04 & 0.08 & 0.08 & 0.19 & 0.18 & 0.25 & 0.15 & 0.23 & 0.13 & \multicolumn{1}{c}{---} & 0.33 & 1 & 0.13 & 1 & 1 & 0\\
\quad \phantom{0}5 & 0.06 & 0.12 & 0.09 & 0.10 & 0.14 & 0.14 & 0.18 & 0.17 & 0.19 &0.10 & \multicolumn{1}{c}{---} & 1 & 1 & 1 & 1 & 1\\
\quad \phantom{0}6 & 0.06 & 0.12 & 0.10 & 0.14 & 0.19 & 0.20 & 0.19 & 0.21 & 0.18 &0.17 & 0.27 & \multicolumn{1}{c}{---} & 1 & 1 & 1 & 1\\
\quad \phantom{0}7 & 0.13 & 0.21 & 0.14 & 0.10 & 0.10 & 0.14 & 0.27 & 0.21 & 0.28 &0.09 & 0.30 & 0.23 & \multicolumn{1}{c}{---} & 1 & 1 & 1\\
\quad \phantom{0}8 & 0.05 & 0.09 & 0.07 & 0.14 & 0.39 & 0.19 & 0.16 & 0.19 & 0.17 &0.15 & 0.26 & 0.26 & 0.20 & \multicolumn{1}{c}{---} & 1 & 0.95\\
\quad \phantom{0}9 & 0.07 & 0.13 & 0.09 & 0.11 & 0.20 & 0.16 & 0.20 & 0.21 & 0.23 &0.12 & 0.31 & 0.25 & 0.30 & 0.42 & \multicolumn{1}{c}{---} & 1\\
\quad 10 & 0.14 & 0.16 & 0.13 & 0.06 & 0.05 & 0.09 & 0.20 & 0.13 & 0.20 &0.05 & 0.19 & 0.12 & 0.32 & 0.11 & 0.20 & \multicolumn{1}{c@{}}{---} \\
\hline
\end{tabular*}
\end{sidewaystable}

\begin{figure}

\includegraphics{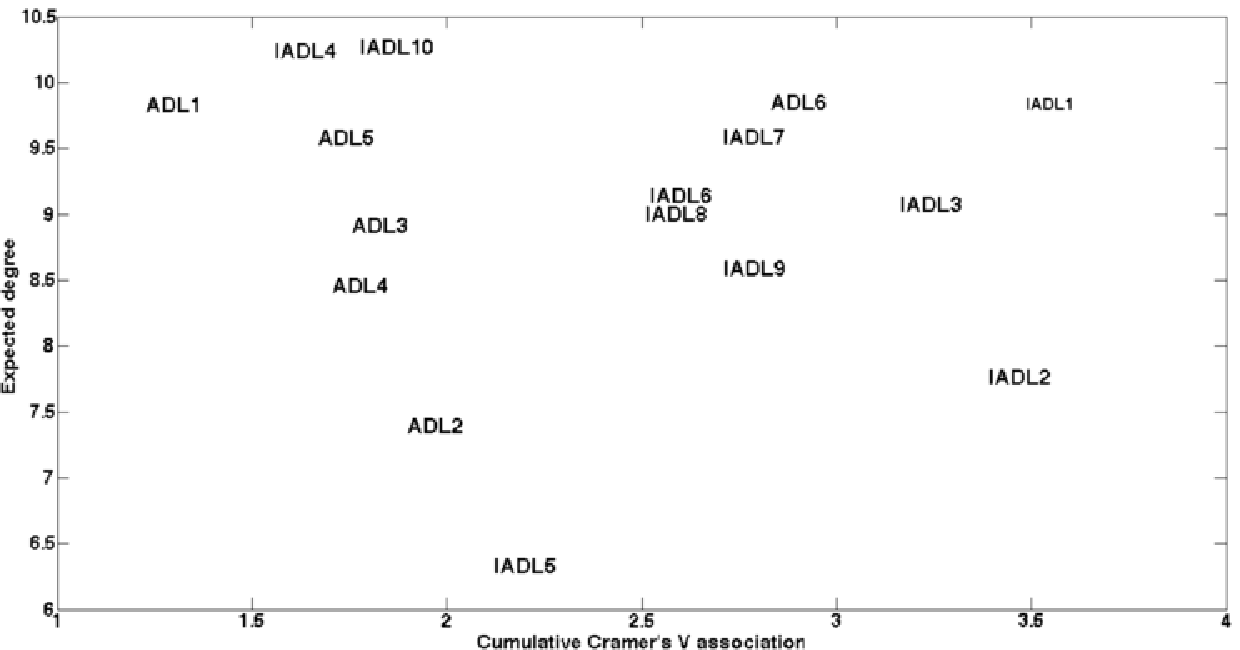}

\caption{Cumulative Cram\'{e}r's V associations (x-axis) and posterior
expected degrees (y-axis) of the $16$ disability measures from the
NLTCS functional disability data.}\label{fig:degreenltcs}
\end{figure}

The topology of the sampled graphs is indicative of the relative
importance of each disability measure with respect to the others in the
latent variables space. The structure of a graph can be summarized by
the number of neighbors of each vertex, that is, the number of edges
that involve each variable. This is usually called the degree of a
vertex. A larger degree indicates an increased number of interactions
in which a latent variable participates. Since in the NLTCS data all
the latent variables are positively associated with each other, having
one disability increases the likelihood of having other disabilities.
The degree of a variable reflects the number of disabilities that are
not conditionally independent of this variable given the
others.\looseness=-1

In the observed variables space we quantify the relative importance of
a~variable $X_{v_{1}}$ as the sum of the Cram\'{e}r's V associations
$\rho_{v_{1},v_{2}}$ between $X_{v_{1}}$ and some other variable
$X_{v_{2}}$. When computing these cumulative Cram\'{e}r's~V associations we
assume that the $120-88=32$ pairwise associations with a~Bayes factor
below $100$ are set to zero. Figure \ref{fig:degreenltcs} shows the
posterior expected degrees of the $16$ disability measures plotted
against the corresponding cumulative Cram\'{e}r's V associations. We see
that IADL4 (cooking) and IADL10 (telephoning) stand out in the latent
space. Most individuals included in the survey ($67.6\%$) are unable to
cook, hence, there is no surprise that IADL4 is the second most
connected variable. However, only a relatively small number of people
($10.6\%$) cannot use the telephone on their own. In fact, more people
are disabled with respect to any of the other $15$ measures. As such,
it might be counterintuitive to see that IADL10 has the highest degree
of connectivity. In the observed variables space the top three
cumulative Cram\'{e}r's~V associations are obtained for IADL1, IADL2 and
IADL3. We note that IADL1 (doing heavy house work) and IADL2 (doing
light house work) are nested, hence, we would expect their association
scores to be related. This indicates a good degree of consistency of
the dependency structure identified by the CGGMs. Since IADL1 is also
highly connected in the latent space, Figure \ref{fig:degreenltcs}
suggests that IADL1 is key to a principled assessment of the disability
level of a person.

The CGGMs clearly show that the $16$ disability measures recorded in
the NLTCS data should not be treated on an equal footing. Some measures
such as IADL1 or IADL10 indicate more serious disabilities than others,
which is not necessarily reflected in the number of people reporting
that particular disability. Simply counting the number of disabilities
a person has can be very misleading when evaluating the overall
disability level of an individual. This remark could shed a new light
on the findings reported in \citet{ manton2gu22001} who only make the
distinction between ADLs and IADLs.

\section{Discussion} \label{sec:discussion}

The inference approach we presented in this paper extends Gaussian
graphical models to data sets in which the multivariate normal
assumption for the observed variables is unlikely to hold. The CGGMs
capture conditional independence relationships among a set of latent
variables that are in a one-to-one relationship with the set of
observed variables. The fact that the number of latent variables
coincides with the number of observed variables avoids the difficult
statistical issue of having to select the number of latent classes---see the excellent discussions in \citet{erosheva2008} and
\citet{fien1hers1rina1zhou12008}.

Our goal was to model dependencies separately from the univariate
margi\-nal distribution of each variable. As such, we did not include a
parametric representation of the marginal distributions in our
framework. \citet{pitt2et22006} give a Bayesian approach to model
conditional independence relationships in Gaussian copulas in which the
univariate marginal distributions are allowed to depend on a set of
parameters and on certain sets of explanatory variables. There is a
definite possibility to combine our prior specification for the
precision matrix for the latent variables with the methods of
\citet{pitt2et22006} into a procedure that takes into account the
uncertainty in the specification of the univariate distributions.

The CGGMs are applicable to any observational study for the purpose of
identifying conditional independence relationships. The only
requirement is that the observed variables are binary, ordinal or
continuous. The extended rank likelihood [\citet{hoff22007}] is a key
component of our framework. A necessary condition for its correct
application is that there exists an ordering of the possible values of
any observed variable---see Section \ref{sec:ordcar}. Our framework
does not allow the presence of discrete variables that are not binary
or ordinal.

Although the interactions among the latent variables do not go beyond
second-order moments, CGGMs give sensible results in the analysis of
sparse contingency tables because they allow inference through Bayesian
model averaging. By contrast, log-linear models contain higher-order
interaction terms but model averaging is no longer an option: the same
interaction term has a different interpretation in various log-linear
models. As such, one has to choose one log-linear model and perform
inference given this single model. When the sample size is small with
respect to the total number of possible models, such a determination
might not be appropriate. The data might not contain enough information
to distinguish between log-linear models that are very close to each
other and have almost the same posterior probability---see, for
example, the analysis of the Rochdale data from
\citet{dobra2massam22009}. Our use of CGGMs does not involve choosing
one particular model, but averaging with respect to many models on the
latent space. We hope that CGGMs will play a significant role in many
quantitative fields of research.

\section*{Acknowledgments}

The authors thank Peter Hoff for useful discussions. The authors are
also grateful to Elena Erosheva who provided the NLTCS data. The
authors thank the Editor and anonymous reviewers for their comments
that improved the quality of this writing.

\begin{supplement}[id=suppA]
\sname{Supplement}
\stitle{C++ implementation of copula Gaussian graphical models}
\slink[doi]{10.1214/10-AOAS397SUPP} 
\slink[url]{http://lib.stat.cmu.edu/aoas/397/supplement.zip}
\sdatatype{.zip}
\sdescription{We provide source code for the
methodology described in this paper. Our program takes advantage of
cluster computing to run several Markov chains in parallel. By using
this code, one can replicate the analyses of the Rochdale data and the
NLTCS functional disability data for which we give sample input files.}
\end{supplement}

\printaddresses

\end{document}